%% file: root.tex
\documentclass[letterpaper, 10 pt, conference]{ieeeconf}  %

\usepackage[utf8]{inputenc}
\usepackage[T1]{fontenc}
\usepackage{tabularx}
\usepackage{booktabs}
\usepackage{cite}                                         %
\providecommand{\tightlist}{\setlength{\itemsep}{0pt}\setlength{\parskip}{0pt}}  %
\IEEEoverridecommandlockouts                              %

\usepackage[dvipsnames]{xcolor}
\usepackage{amssymb}
\usepackage{graphicx}
\usepackage{balance}   %

\newcounter{principle}
\newenvironment{principle}[1][]{%
  \par\medskip\noindent
  \refstepcounter{principle}%
  \textbf{Principle~\theprinciple}%
  \if\relax\detokenize{#1}\relax\else\space(#1)\fi%
  \textbf{.}\space\itshape\ignorespaces
}{%
  \par%
}

\title{%
{\normalfont\normalsize\itshape Accepted as a Late Breaking Report (LBR) at IEEE RO-MAN 2026\\
[1.8em]}%
\LARGE \bf
Toward Personalized Social Robots for Child Well-being: Data Requirement Principles from a Recommender-System Perspective
}

\author{Jin Huang$^{1,*}$, Eric Nichols$^{2,*}$, Fethiye Irmak Do\u{g}an$^{1}$, and Hatice Gunes$^{1}$%
\thanks{$^{*}$These authors contributed equally and are co-first authors.}%
\thanks{$^{1}$J. Huang, F. I. Do\u{g}an, and H. Gunes are with the University of
        Cambridge, Cambridge, United Kingdom
        {\tt\small \{jh2642, fid21, hg410\}@cam.ac.uk}}%
\thanks{$^{2}$E. Nichols is with Honda Research Institute Japan, Wako, Japan
        {\tt\small e.nichols@jp.honda-ri.com}}%
\thanks{The work of J. Huang was supported by the EU's Horizon Europe research and innovation programme under the Marie Sk\l{}odowska-Curie Actions Postdoctoral Fellowships (European Fellowship) 2024, grant agreement no.~101203728 --- SOCIALADAPT --- HORIZON-MSCA-2024-PF-01. The works of F. I. Do\u{g}an \& H. Gunes have been supported by CHANSE \& NORFACE through the MICRO project, funded by ESRC/UKRI grant ref.~UKRI572.  Views and opinions expressed are those of the author(s) only and do not necessarily reflect those of the funding bodies. Neither the EU nor the granting authorities can be held responsible. 
\textbf{Contributions:} Conceptualization: JH, EN. Funding acquisition: JH, HG. Methodology: JH, EN. Writing, Review \& Editing: JH, EN, FID, HG. Supervision \& project administration: JH, HG.}
}

\begin{document}

\maketitle
\thispagestyle{empty}
\pagestyle{empty}

\begin{abstract}

Social robots are increasingly deployed in clinical settings to support the well-being of children, where effective support must be personalized to each child.
Personalization, choosing the robot action best suited to each child, can be framed as a recommendation problem, and a recently proposed recommender-system framework for social robots offers a principled approach through user profiling, ranking, and responsible computing.
Instantiating it, however, is blocked not by the model but by the data, which is hard to gather.
A child's state shifts within and across visits, so no fixed description of the user holds.
Within a session, the few signals of whether the robot's actions helped are weak and indirect.
Across sessions, children are rarely seen more than once, and anonymization breaks the identity needed to link visits.
Because care cannot be randomized, existing data is observational, biased toward whatever was already done.
Each is a familiar recommender-system problem, and we propose four data principles in response: an integrated profile, effectiveness signals, linkable coverage, and an exposure record logged at collection time.
We identify which of these principles each capability requires, and frame them as concrete guidelines for data collection.

\end{abstract}

\input{Sections/sec-intro}
\input{Sections/sec-background}
\input{Sections/sec-existing}

\input{Sections/sec-data}
\input{Sections/sec-minimal}
\input{Sections/sec-con}

\addtolength{\textheight}{-12cm}   %

\balance
\bibliographystyle{IEEEtran}
\bibliography{references_shorten}

\end{document}

%% file: Sections/sec-intro.tex
\section{Introduction}
\label{sec:intro}

\begin{figure*}[!t]
\centering
\includegraphics[width=0.85\textwidth]{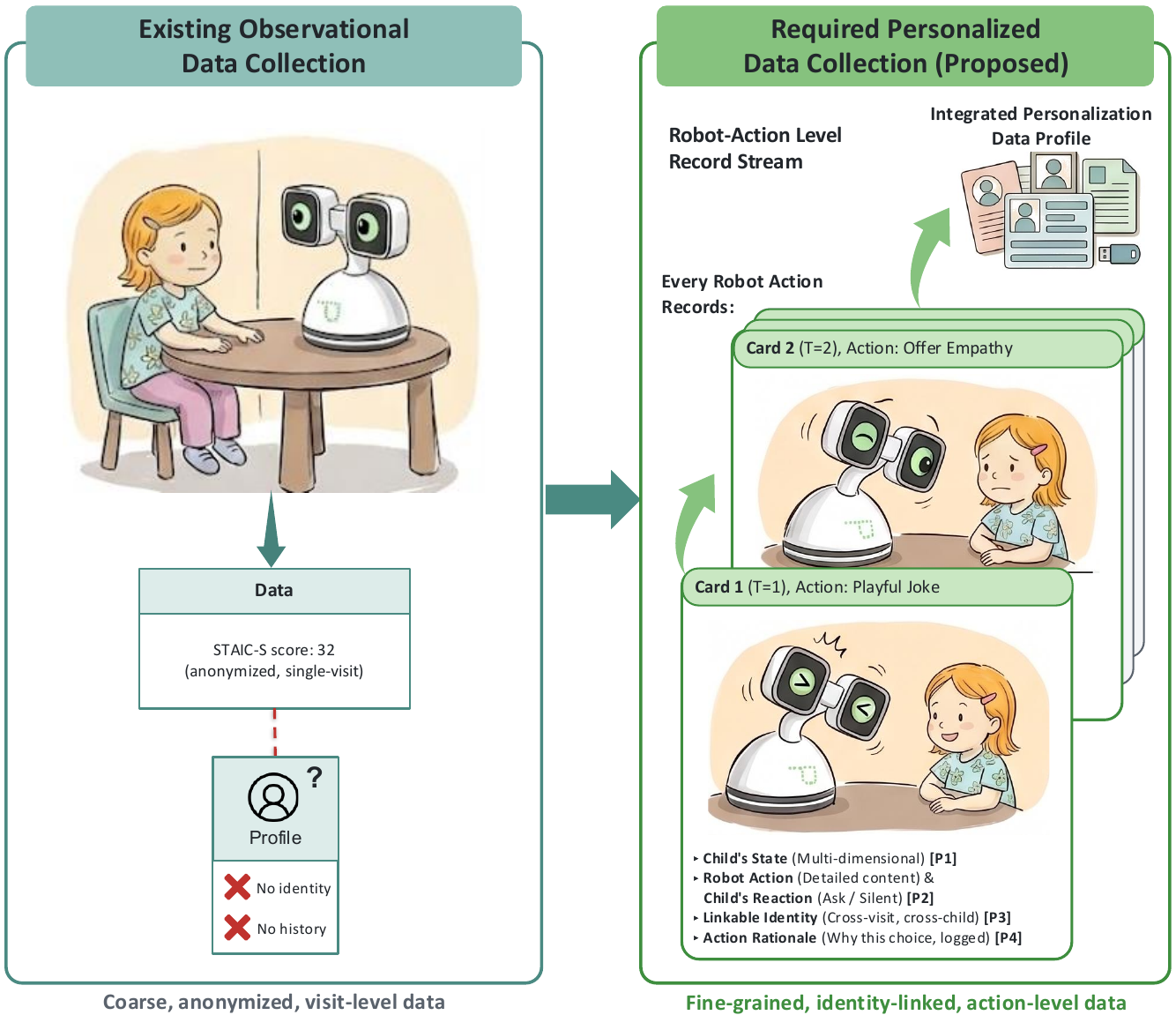} \vspace{-5pt}
\caption{Existing vs.\ required data collection for personalized social robots in child well-being settings. Left: existing robot-administered assessments reduce a session to a single anonymized score, which can neither build a user profile nor link to one. Right: the proposed collection logs one record per robot action, holding the child's state (P1), the action and the reaction that follows (P2), and why the action was chosen (P4), all tied to a linkable identity across visits and children (P3) into an integrated profile that recommender-system models can learn from. P1--P4 refer to the four data principles of Section~\ref{sec:data}. The robot depicted is the social robot Haru~\cite{gomez2018haru}.}
\label{fig:data-collection} \vspace{-5pt}
\end{figure*}

Ensuring the well-being of children is a priority shared by parents, educators, and clinicians, and it depends not only on a child's physical care but on how the people around them respond to their needs. This is most acute in clinical settings, where illness and medical procedures place children under stress. Social robots are increasingly brought in to offer companionship and emotional reassurance, and to help children understand what is happening to them~\cite{rossi22, nigro2025radiological, blansonhenkemans2013diabetes}. However, children vary widely in developmental stage, anxiety, prior knowledge, and language, so effective support must be personalized to each child~\cite{smakman21, dehaas24}.

This need is concrete in real-world applications. For instance, at Hospital Virgen del Rocío in Seville, Spain, the social robot Haru~\cite{gomez2018haru} provides companionship and conducts anxiety screening for children undergoing cancer treatment~\cite{nichols2026anxiety}. Many of these children, such as those with leukemia, return repeatedly over months of care, and their anxiety, understanding, and emotional state differ from one child to the next and shift from visit to visit. To effectively support them, the robot must adapt both to the individual child and to how that child changes over time. Personalization is, therefore, the capability these applications require.

Yet the data needed to learn it is hard to gather. We identify four challenges:

\begin{enumerate}
\def\labelenumi{(\roman{enumi})}
\tightlist
\item
  A child's anxiety, attention, and understanding shift within and across visits, so no fixed description of the user holds~\cite{leite2014empathic}.
\item
  Within a session, children typically give few direct signals of whether the interaction helped, often at its conclusion, and the rest must be inferred from weak, indirect cues~\cite{castellano2009detecting}.
\item
  Across sessions, most children are seen only once while a few return repeatedly, and consent and anonymization break the per-child identity that linking visits depends on, so a usable cross-session history rarely exists~\cite{kabacinska20, hintze2018pseudonymisation}.
\item
  Because care cannot be randomized, existing data is observational, biased toward whatever was already done~\cite{kyriacou2016confounding}.
\end{enumerate}

These are not limitations of scale: collecting more data under the same designs would not yield what personalization requires; the data must be designed, not merely accumulated.

None of these challenges is new. Each is a problem recommender systems are built to handle: a non-stationary user state (i), implicit feedback (ii), cold-start and unlinkable histories (iii), and off-policy bias (iv). Building on a recent recommender-system framework for social robots~\cite{huang2026reimagining}, we argue that what stands between that framework and these applications is not the model but the data it can learn from.

We therefore propose four data principles, one per challenge (see Fig.~\ref{fig:data-collection}): (i) an integrated profile that tracks the changing child, (ii) effectiveness signals that capture implicit feedback, (iii)~linkable coverage that connects histories across children and visits, and (iv) an exposure record that corrects for non-randomized care. The rest of the paper develops them, examines how far existing data meets them, and identifies which principles each capability requires.

%% file: Sections/sec-background.tex
\begin{table*}[!htp]
\centering
\small
\renewcommand{\arraystretch}{1.25}
\caption{Representative robot-administered assessment and patient-communication studies, organized by data challenges (i)--(iii); challenge~(iv), exposure, is a property of collection design rather than a per-study column.}
\vspace{-5pt}
\label{tab:gaps}
\begin{tabular}{@{}l p{1.8cm} p{4.5cm} p{2.6cm} c c@{}}
\toprule
&  & User data (i) & Feedback (ii) & \multicolumn{2}{c}{Sample (iii)} \\
\cmidrule(lr){3-3}\cmidrule(lr){4-4}\cmidrule(l){5-6}
Study & Robot & Instrument / construct & Feedback unit & $N$ & Sessions \\
\midrule
Briggs et al.\ 2015~\cite{briggs15}            & NAO             & PDQ-39 (Parkinson's health status)                    & Session-level scores       & 13                          & single \\
Van der Putte et al.\ 2019~\cite{vdputte19}    & Pepper          & Inpatient questionnaires (history, pain, sleep, etc.) & Session-level scores       & 35                          & single \\
Smakman et al.\ 2021~\cite{smakman21}          & NAO             & Age band; prior experience                            & Phase-of-session (VAS-A, FLACC) & 158                    & single \\
Abbasi et al.\ 2022~\cite{abbasi22}            & NAO             & SMFQ, RCADS (child well-being)                         & Session-level scores       & 28                          & single \\
Kaptein et al.\ 2021~\cite{kaptein22}          & NAO + avatar    & Diabetes self-management goals                        & Session-level logs         & 48                          & up to 6 months \\
Abbasi et al.\ 2025~\cite{abbasi25}            & NAO             & SMFQ (child well-being), online                        & Session-level scores       & 40                          & 3 \\
de Araujo et al.\ 2026~\cite{dearaujo26}       & Robios          & GDS-15 (depression, older adults)                     & Session-level score        & 20 & single \\
Nichols et al.\ 2026~\cite{nichols2026anxiety} & Haru            & STAIC-S (state anxiety, pediatric oncology)           & Session-level score        & 81 & single \\
van 't Klooster et al.\ 2026~\cite{vtk26}      & Furhat + GPT-4o & Condition (osteoarthritis)                            & Session-level (UEQ + interview) & 21                  & single \\
\bottomrule
\end{tabular}

\vspace{-5pt}
\end{table*}

\section{Background and Related Work}

\subsection{Social robots and data collection in hospital settings}

Social robots are increasingly deployed in hospitals and clinics for patient-facing roles, including companionship and emotional support~\cite{rossi22}, patient education and explanation~\cite{boumans23}, screening and assessment~\cite{briggs15, abbasi22, abbasi24, dearaujo26}, procedural distraction~\cite{smakman21, dehaas24}, and longer-term self-management and monitoring~\cite{kaptein22, basch16}, across adult and pediatric populations and platforms such as NAO, Pepper, Furhat, and Haru~\cite{charisi20, nichols2026anxiety}. Our concern here is less the breadth of these applications than how they collect data. As Table~\ref{tab:gaps} illustrates, user data comes predominantly from robot-administered standardized instruments and patient-reported outcome measures~\cite{briggs15, vdputte19, abbasi22}; interaction is typically rule-based or scripted, with large language models only recently introduced~\cite{vtk26, abbasi25vlm}; and study designs are mostly single-session, recording outcomes as end-of-session scores and user-experience ratings~\cite{kabacinska20}.

\subsection{Review of RS-inspired framework for social robots}

We build on the framework of~\cite{huang2026reimagining}, which reframes personalization in social robots as a recommendation problem and integrates three modular, plug-and-play components into the robot's cognition pipeline. \emph{User profiling} models a user's preferences at complementary granularities: long-term interests via collaborative filtering, short-term intent via sequential modeling, and fine-grained preferences via knowledge-enhanced modeling. \emph{Ranking} scores and orders candidate robot actions for the current user and selects the most appropriate one to execute. \emph{Responsible computing} constrains these operations to preserve privacy, mitigate bias, and promote fairness.
Instantiating these components in the clinical child well-being setting presupposes data of a particular kind (e.g., data linking each user's profile and state to the robot actions they receive and the responses these elicit); whether it is available in existing datasets or under current data-collection strategies is an open question.

%% file: Sections/sec-existing.tex
\section{How Existing Data Meets Personalization}

Patient-facing social-robot studies, such as those in Table~\ref{tab:gaps}, are built to screen or assess a single construct: the robot administers a standardized instrument, and what the session yields is a single score summarizing the user's state on that construct. 
Personalization needs data of a different kind, a record of how each user responded to what the robot did, linked to that user's state and accumulated across many users.
Against this record, existing data falls short on each of the four challenges of Section~\ref{sec:intro}.

Across these studies, the user is described and recorded through the instrument administered and the single construct it measures, one description per session. The description is fixed in two senses: fixed in advance to the one dimension the instrument targets, although what shifts in the child spans several dimensions at once; and fixed over the interaction, although that shifting state is what the robot's actions must adapt to (\textbf{Challenge~i}).
Feedback takes the form of a questionnaire given once the session is over, a rating of how satisfied the user was with the interaction or the robot overall. It reports whether the session as a whole landed well, but not which of the robot's actions within it the user responded to and which they did not, so the per-action signal a preference model would be trained and evaluated on is absent (\textbf{Challenge~ii}).
Most children are seen only once, and consent and anonymization strip the per-child identity that ties one visit to the next, so a child's sessions cannot be assembled into a history. The per-user history that personalizing to a returning child would build on therefore does not exist (\textbf{Challenge~iii}).
This last challenge is not a shortfall of the existing studies: with only a few conditions to compare, collection can randomize over them and stay unbiased. The per-action preferences personalization needs, however, range over an action space too large to randomize over, so the data is observational, biased toward whatever was already done (\textbf{Challenge~iv}).

%% file: Sections/sec-data.tex
\section{Data Requirements}\label{sec:data}

This section introduces the requirements of data that suit the RS framework
as four principles, one per challenge: an integrated profile (Challenge~i),
effectiveness signals (Challenge~ii), linkable, representative coverage
(Challenge~iii), and an exposure record (Challenge~iv).

\begin{principle}[Integrated profile]\label{prin:profile}
The data should record each user's dimensions jointly as a single multi-layer profile (identity, dynamic emotional state, interaction history, situational context, and multimodal signals), rather than as isolated attributes.
\end{principle}

Existing datasets each model a single construct, because they are built for screening or assessment rather than for personalizing actions, so few datasets hold these dimensions together for the same user.
Yet what action suits a user depends on their combination, not on any attribute alone.
Unless the dimensions are recorded jointly, this combination is never observed, and personalization can respond to only one factor at a time.
For example, a child returning for treatment over months may arrive anxious, yet already familiar with the procedure ahead. Anxiety alone cannot determine the action: comforting and explaining are both reasonable responses to an anxious child. Her familiarity is what decides: since she already knows what will happen, explaining would not help her anxiety, and the robot should comfort her instead.

\begin{principle}[Effectiveness signals]\label{prin:signal}
The data should align each robot action, identified by its content rather than as an opaque item, with the user behavior that follows it, both explicit behavior (follow-up questions, requests to rephrase, a brief per-action rating) and implicit signals (hesitations, gaze, facial expression), so that fine-grained preference for individual actions can be inferred and generalized to new ones.
\end{principle}

Recovering the per-action signal does not require users to report on each action, which would be impractical. The implicit signals and the explicit behaviors that arise unprompted are logged passively from the interaction stream and aligned to whichever robot action prompted them, without extra effort from the user.
Combining explicit and implicit feedback in this way has been shown to
improve activity personalization in assistive social robots~\cite{maroto2025personalizing}.
For example, when the robot explains an upcoming procedure to the child, whether the child asks a follow-up question or falls silent and looks away is a signal tied to that explanation, which a single `end-of-visit questionnaire' cannot recover.

\begin{principle}[Linkable, representative coverage]\label{prin:coverage}
Rather than rely on the long per-user histories that clinical settings rarely afford, the data should link each user's interactions to a stable, extensible profile, and span multiple sites and the demographic range of clinical settings, so that a user with a sparse history can be served by drawing on similar users, while the same profile lets personalization deepen as interactions accumulate.
\end{principle}

Personalizing to a user over time is often approached by accumulating a longitudinal record of the same user across sessions~\cite{clabaugh19}, which clinical settings, with their brief or one-off encounters, rarely afford. Recommender systems offer a different route: collaborative filtering and related methods are built to work from sparse histories, transferring preferences from similar users to one whose own record is thin. Personalization, therefore, does not stall when individual histories are short and deepens as they grow.
For example, most children in a hospital ward are seen only once, while a child with leukemia returns over months for care; a linkable profile lets the robot serve a child visiting for the first time by drawing on similar children, and enables personalization for returning children to improve with each visit.

\begin{principle}[Exposure logging]\label{prin:exposure}
Instead of the coarse preferences targeted by existing datasets, which can be studied by randomizing over a few conditions, the data should capture fine-grained preferences over which robot action suits which user; but the space of actions is too large to randomize over, so such data will be observational and biased, where a robot action may appear effective only because it was shown to users likely to respond well. The data should therefore record, at collection time, why each robot action was given to each user, so that an action's effect can be separated from the reasons it was shown.
\end{principle}

Exposure cannot be reconstructed retrospectively. What the system is built on when selecting an action, e.g., the inferred user profile, the candidate robot actions, and the selection policy, exists only at the moment of selection; once the interaction is over, it cannot be recovered. Exposure must thus be recorded as collection happens, not inferred afterward.
With it in place, the framework's responsible-computing component can apply standard debiasing, for example, inverse-propensity weighting~\cite{ips}, to separate an action's effect from the reasons it was selected.
For example, if the robot comforts mainly the most anxious children, the comforted children will still look worse afterward, so the outcomes suggest that comforting failed when it may in fact have helped. The impression can be corrected only if the data records why comfort was chosen, namely that the child appeared anxious, so that comforted children are compared with equally anxious ones.

Together, these four principles specify what the RS framework needs from real data.
Table~\ref{tab:req} maps the four onto the framework's components: each principle cuts across several rather than serving one, so the data cannot be assembled piecemeal.

\begin{table}[t]
\centering
\small
\renewcommand{\arraystretch}{1.3}
\caption{Data requirements and the framework components that consume them.}
\vspace{-8pt}
\label{tab:req}
\begin{tabular}{@{}l c c c@{}}
\toprule
Principles & UP & R & RC \\
\midrule
Integrated profile (i)                  & $\checkmark\checkmark$ & $\checkmark$ & $\checkmark$ \\
Effectiveness signals (ii)              & $\checkmark\checkmark$ & $\checkmark\checkmark$ &  \\
Linkable, representative coverage (iii) & $\checkmark\checkmark$ & $\checkmark$ & $\checkmark\checkmark$ \\
Exposure record (iv)                    &  & $\checkmark$ & $\checkmark\checkmark$ \\
\bottomrule
\end{tabular}

\vspace{3pt}
{\footnotesize
UP: user profiling; R: ranking; RC: responsible computing. \\ $\checkmark\checkmark$: primary consumer; $\checkmark$: also relies on.
}
\vspace{-10pt}
\end{table}

%% file: Sections/sec-minimal.tex
\section{Minimal Required Principles per Capability}

While the four principles describe the data the framework would ideally draw on, running it does not require all of them: the data it needs depends on the targeted capability.

Two of the principles are not tied to any particular capability but underlie all of them.
An integrated user profile (Principle~\ref{prin:profile}) supports personalization in any capability, but how complete it can be is flexible: richer profile dimensions improve modeling where privacy permits, while in settings such as hospitals where they cannot be recorded, the framework still models individual preference from interaction history.
The exposure record (Principle~\ref{prin:exposure}) is likewise required whenever actions are personalized rather than assigned at random. 

The capabilities below differ in what they additionally require:
\textbf{Within-session personalization} models a user and selects actions within a single visit. It additionally needs per-action feedback and an action representation (Principle~\ref{prin:signal}) across a population of users, so that collaborative filtering can draw on similar users to compensate for a sparse per-user signal.
\textbf{Cross-session personalization} tracks a user across return visits, allowing both long-term preference and short-term state to be modeled. It additionally needs a stable identifier linking the user's sessions (Principle~\ref{prin:coverage}).
\textbf{Cold start} models a user with no prior interaction to draw on. One option is to elicit preference directly: a few probing actions, such as briefly asking the user, yield follow-up responses (Principle~\ref{prin:signal}) that quickly establish an initial sense of the user's preferences.
\textbf{Cross-domain transfer} reuses preferences learned in one setting in another, such as from chitchat to anxiety screening. It requires that the per-action responses (Principle~\ref{prin:signal}) be tagged with the situational context they occur in (Principle~\ref{prin:profile}), so that preferences can be attributed to the right setting and carried across.

%% file: Sections/sec-con.tex
\section{Discussion and Conclusion}
In this paper, we presented dataset challenges for personalization in robot-mediated child well-being applications from a recommender-system perspective. We note that
two practical issues remain open: how to collect the data, and how much is enough. Collection is constrained by consent and anonymization, which limit what can be recorded and linked. This is not fatal: where a fuller profile is unavailable, the framework still personalizes from interaction history alone. What matters is therefore not whether the principles are fully met, but how the data is gathered in practice, for instance, through privacy-preserving or federated collection; the less data is collected, the harder personalization becomes.
How much suffices has no fixed answer, and is empirical along three axes: \emph{scale}, how many users before a thin history can be served by transfer from similar ones; \emph{diversity}, how many sites and how wide a demographic range before preferences carry across populations; and \emph{temporal depth}, how many return visits before a changing user state can be modeled rather than sampled once.

The four principles specify what data personalization needs, and which subset of them each capability requires; existing data supplies only fragments. Whether social robots can personalize their interactions, therefore, turns less on new models than on collecting the right data from the start.

%% file: root.bbl
\begin{thebibliography}{10}
\providecommand{\url}[1]{#1}
\csname url@samestyle\endcsname
\providecommand{\newblock}{\relax}
\providecommand{\bibinfo}[2]{#2}
\providecommand{\BIBentrySTDinterwordspacing}{\spaceskip=0pt\relax}
\providecommand{\BIBentryALTinterwordstretchfactor}{4}
\providecommand{\BIBentryALTinterwordspacing}{\spaceskip=\fontdimen2\font plus
\BIBentryALTinterwordstretchfactor\fontdimen3\font minus
  \fontdimen4\font\relax}
\providecommand{\BIBforeignlanguage}[2]{{%
\expandafter\ifx\csname l@#1\endcsname\relax
\typeout{** WARNING: IEEEtran.bst: No hyphenation pattern has been}%
\typeout{** loaded for the language `#1'. Using the pattern for}%
\typeout{** the default language instead.}%
\else
\language=\csname l@#1\endcsname
\fi
#2}}
\providecommand{\BIBdecl}{\relax}
\BIBdecl

\bibitem{gomez2018haru}
R.~Gomez, D.~Szapiro, K.~Galindo, and K.~Nakamura, ``Haru: Hardware design of
  an experimental tabletop robot assistant,'' in \emph{Proc. ACM/IEEE Int.
  Conf. Hum.-Robot Interact. (HRI)}, 2018, pp. 233--240.

\bibitem{rossi22}
S.~Rossi, S.~J. Santini, D.~Di~Genova, G.~Maggi, A.~Verrotti, G.~Farello,
  R.~Romualdi, A.~Alisi, A.~E. Tozzi, and C.~Balsano, ``Using the social robot
  {NAO} for emotional support to children at a pediatric emergency department:
  Randomized clinical trial,'' \emph{J. Med. Internet Res.}, vol.~24, no.~1, p.
  e29656, 2022.

\bibitem{nigro2025radiological}
M.~Nigro, A.~Righini, and M.~Spitale, ``Exploring the use of social robots to
  prepare children for radiological procedures: A focus group study,'' in
  \emph{Proc. IEEE Int. Conf. Robot Hum. Interact. Commun. (RO-MAN)}, 2025.

\bibitem{blansonhenkemans2013diabetes}
O.~A. Blanson~Henkemans, B.~P.~B. Bierman, J.~Janssen, M.~A. Neerincx,
  R.~Looije, H.~van~der Bosch, and J.~A.~M. van~der Giessen, ``Using a robot to
  personalise health education for children with diabetes type 1: A pilot
  study,'' \emph{Patient Educ. Couns.}, vol.~92, no.~2, pp. 174--181, 2013.

\bibitem{smakman21}
M.~H.~J. Smakman, K.~Smit, L.~Buser, T.~Monshouwer, N.~van Putten, T.~Trip,
  C.~Schoof, D.~F. Preciado, E.~A. Konijn, E.~M. van~der Roest, and W.~M.
  Tiel~Groenestege, ``Mitigating {Children's} pain and anxiety during blood
  draw using social robots,'' \emph{Electronics}, vol.~10, no.~10, p. 1211,
  2021.

\bibitem{dehaas24}
M.~De~Haas, K.~Smit, D.~F. Preciado~Vanegas, E.~van~der Roest, M.~Smakman, and
  W.~Tiel~Groenestege, ``The effect of a social robot on {Children's} pain and
  anxiety during blood draw,'' in \emph{Proc. ACM Interact. Des. Child. Conf.
  (IDC)}, 2024, pp. 776--780.

\bibitem{nichols2026anxiety}
E.~Nichols, N.~P{\'e}rez~Higueras, M.~Orozco, G.~P{\'e}rez,
  G.~{\'A}lvarez-Benito, J.~G. Amores-Carredano, L.~Merino, and R.~Gomez,
  ``Autonomous robot-administered anxiety screening in pediatric oncology,'' in
  \emph{Proc. IEEE Int. Conf. Robot Hum. Interact. Commun. (RO-MAN)}, 2026, to
  appear.

\bibitem{leite2014empathic}
I.~Leite, G.~Castellano, A.~Pereira, C.~Martinho, and A.~Paiva, ``Empathic
  robots for long-term interaction,'' \emph{Int. J. Soc. Robot.}, vol.~6,
  no.~3, pp. 329--341, 2014.

\bibitem{castellano2009detecting}
G.~Castellano, A.~Pereira, I.~Leite, A.~Paiva, and P.~W. McOwan, ``Detecting
  user engagement with a robot companion using task and social
  interaction-based features,'' in \emph{Proc. Int. Conf. Multimodal Interfaces
  (ICMI-MLMI)}, 2009, pp. 119--126.

\bibitem{kabacinska20}
K.~Kabaci{\'n}ska, T.~J. Prescott, and J.~M. Robillard, ``Socially assistive
  robots as mental health interventions for children: A scoping review,''
  \emph{Int. J. Soc. Robot.}, vol.~13, no.~5, pp. 919--935, 2020.

\bibitem{hintze2018pseudonymisation}
M.~Hintze and K.~El~Emam, ``Comparing the benefits of pseudonymisation and
  anonymisation under the {GDPR},'' \emph{J. Data Prot. Priv.}, vol.~2, no.~2,
  pp. 145--158, 2018.

\bibitem{kyriacou2016confounding}
D.~N. Kyriacou and R.~J. Lewis, ``Confounding by indication in clinical
  research,'' \emph{JAMA}, vol. 316, no.~17, pp. 1818--1819, 2016.

\bibitem{huang2026reimagining}
J.~Huang, F.~I. Do{\u{g}}an, and H.~Gunes, ``Reimagining social robots as
  recommender systems: Foundations, framework, and applications,'' in
  \emph{Proc. ACM/IEEE Int. Conf. Hum.-Robot Interact. (HRI)}, 2026, pp.
  406--416.

\bibitem{briggs15}
P.~Briggs, M.~Scheutz, and L.~Tickle-Degnen, ``Are robots ready for
  administering health status surveys?: First results from an {HRI} study with
  subjects with {Parkinson's} disease,'' in \emph{Proc. ACM/IEEE Int. Conf.
  Hum.-Robot Interact. (HRI)}, 2015, pp. 327--334.

\bibitem{vdputte19}
D.~van~der Putte, R.~Boumans, M.~Neerincx, M.~O. Rikkert, and M.~de~Mul, ``A
  social robot for autonomous health data acquisition among hospitalized
  patients: An exploratory field study,'' in \emph{Proc. ACM/IEEE Int. Conf.
  Hum.-Robot Interact. (HRI)}, 2019, pp. 658--659.

\bibitem{abbasi22}
N.~I. Abbasi, M.~Spitale, J.~Anderson, T.~Ford, P.~B. Jones, and H.~Gunes,
  ``Can robots help in the evaluation of mental wellbeing in children? {An}
  empirical study,'' in \emph{Proc. IEEE Int. Conf. Robot Hum. Interact.
  Commun. (RO-MAN)}, 2022, pp. 1459--1466.

\bibitem{kaptein22}
F.~Kaptein, B.~Kiefer, A.~Cully, O.~Celiktutan, B.~Bierman,
  R.~Rijgersberg-Peters, J.~Broekens, W.~van Vught, M.~van Bekkum, Y.~Demiris,
  and M.~A. Neerincx, ``A cloud-based robot system for long-term interaction:
  Principles, implementation, lessons learned,'' \emph{ACM Trans. Hum.-Robot
  Interact.}, vol.~11, no.~1, pp. 1--27, 2021.

\bibitem{abbasi25}
N.~I. Abbasi, G.~Laban, T.~Ford, P.~B. Jones, and H.~Gunes, ``A longitudinal
  study of child wellbeing assessment via online interactions with a social
  robot,'' \emph{ACM Trans. Hum.-Robot Interact.}, vol.~14, no.~3, pp. 1--35,
  2025.

\bibitem{dearaujo26}
B.~S. de~Araujo, M.~Fantinato, M.~Cachioni, M.~S. Yassuda, R.~C. de~Melo, S.~M.
  Peres, and P.~C.~K. Hung, ``Exploring the use of social robots for depression
  screening among older adults: An expanded feasibility study,'' \emph{IEEE
  Access}, vol.~14, pp. 22\,977--22\,997, 2026.

\bibitem{vtk26}
J.-W. J.~R. van~'t Klooster, M.~Capasso, D.~van Gorssel, E.~Vrolijk,
  G.~Rettagliata, D.~Gerritsen, M.~Hegeman, E.~Tauro, E.~G. Caiani, and H.~E.
  Vonkeman, ``A {GPT}-reinforced social robot for patient communication: a
  pilot study,'' \emph{Front. Digit. Health}, vol.~7, 2026.

\bibitem{boumans23}
R.~Boumans, R.~Melis, T.~Bosse, and S.~Thill, ``A social robot for explaining
  medical tests and procedures: An exploratory study in the wild,'' in
  \emph{Companion ACM/IEEE Int. Conf. Hum.-Robot Interact. (HRI)}, 2023, pp.
  263--267.

\bibitem{abbasi24}
N.~I. Abbasi, G.~Laban, T.~Ford, P.~B. Jones, and H.~Gunes, ``Robotising
  psychometrics: Validating wellbeing assessment tools in child-robot
  interactions,'' in \emph{Proc. IEEE Int. Conf. Robot Hum. Interact. Commun.
  (RO-MAN)}, 2024, pp. 1651--1658.

\bibitem{basch16}
E.~Basch, A.~M. Deal, M.~G. Kris, H.~I. Scher, C.~A. Hudis, P.~Sabbatini,
  L.~Rogak, A.~V. Bennett, A.~C. Dueck, T.~M. Atkinson, J.~F. Chou, D.~Dulko,
  L.~Sit, A.~Barz, P.~Novotny, M.~Fruscione, J.~A. Sloan, and D.~Schrag,
  ``Symptom monitoring with patient-reported outcomes during routine cancer
  treatment: A randomized controlled trial,'' \emph{J. Clin. Oncol.}, vol.~34,
  no.~6, pp. 557--565, 2016.

\bibitem{charisi20}
V.~Charisi, E.~Gomez, G.~Mier, L.~Merino, and R.~Gomez, ``Child-robot
  collaborative problem-solving and the importance of child's voluntary
  interaction: A developmental perspective,'' \emph{Front. Robot. AI}, vol.~7,
  p.~15, 2020.

\bibitem{abbasi25vlm}
N.~I. Abbasi, F.~I. Do{\u{g}}an, G.~Laban, J.~Anderson, T.~Ford, P.~B. Jones,
  and H.~Gunes, ``Robot-led vision language model wellbeing assessment of
  children,'' in \emph{Proc. IEEE Int. Conf. Robot Hum. Interact. Commun.
  (RO-MAN)}, 2025, pp. 59--64.

\bibitem{maroto2025personalizing}
M.~Maroto-G{\'o}mez, M.~Malfaz, J.~C. Castillo, {\'A}.~Castro-Gonz{\'a}lez, and
  M.~{\'A}. Salichs, ``Personalizing activity selection in assistive social
  robots from explicit and implicit user feedback,'' \emph{Int. J. Soc.
  Robot.}, vol.~17, no.~10, pp. 1999--2017, 2025.

\bibitem{clabaugh19}
C.~Clabaugh, K.~Mahajan, S.~Jain, R.~Pakkar, D.~Becerra, Z.~Shi, E.~Deng,
  R.~Lee, G.~Ragusa, and M.~Matari{\'c}, ``Long-term personalization of an
  in-home socially assistive robot for children with autism spectrum
  disorders,'' \emph{Front. Robot. AI}, vol.~6, p. 110, 2019.

\bibitem{ips}
T.~Schnabel, A.~Swaminathan, A.~Singh, N.~Chandak, and T.~Joachims,
  ``Recommendations as treatments: Debiasing learning and evaluation,'' in
  \emph{Proc. Int. Conf. Mach. Learn. (ICML)}, ser. Proc. Mach. Learn. Res.
  (PMLR), vol.~48, 2016, pp. 1670--1679.

\end{thebibliography}
